\documentclass[journal]{IEEEtran}
\usepackage{amsmath, amssymb, amsfonts}
\usepackage{graphicx}
\usepackage{algorithm}
\usepackage{algpseudocode}
\usepackage{cite}
\usepackage{url}
\usepackage{amsthm}
\usepackage{balance}
\usepackage{xcolor}
\usepackage{tikz}
\usetikzlibrary{shapes, arrows, positioning, calc}

\newtheorem{definition}{Definition}

\begin{document}
\title{A Path-Survival Analytical Framework for SCL Decoding of Polar Codes}
\author{Xianbin~Wang,~Zhichao~Liu,~Yuan~Li,~Huazi~Zhang,~Jiajie~Tong,~Jun~Wang,~and~Wen~Tong%
        \thanks{X. Wang, Z. Liu, Y. Li, H. Zhang, J. Tong, J. Wang, and W. Tong are with Huawei Technologies Co., Ltd., Shenzhen, China (e-mail: \{wangxianbin1, liuzhichao28, liyuan299, zhanghuazi, tongjiajie, justin.wangjun, tongwen\}@huawei.com).}
}
\maketitle
\begin{abstract}
A theoretical analysis of CRC-aided successive cancellation list (CA-SCL) decoding for polar codes remains an open problem, despite its widespread practical adoption. While low-density parity-check (LDPC) codes benefit from mature analytical tools, such as density evolution (DE), for predicting the performance of belief-propagation (BP) decoding, similar techniques are not directly applicable to CA-SCL decoding. This limitation stems from the complex path-pruning mechanism inherent in CA-SCL decoding. In this paper, we propose an analytical framework based on a novel path-survival model that captures the evolution of the correct path's rank during decoding. The proposed framework enables efficient prediction of CA-SCL decoding performance without requiring exhaustive list-specific Monte Carlo simulations. Extensive numerical evaluations demonstrate its effectiveness across a wide range of code lengths, code rates, list sizes, and channel models.
\end{abstract}
\begin{IEEEkeywords}
	Polar codes, successive cancellation list decoding, performance analysis.
\end{IEEEkeywords}

\section{Introduction}
\label{sec:introduction}
Decoding performance analysis is fundamental to the design and optimization of channel codes. It not only enables performance prediction without resorting to extensive Monte Carlo simulations, but also provides theoretical insights into algorithmic behavior that facilitate decoder enhancement. For low-density parity-check (LDPC) codes, analytical tools such as density evolution~\cite{DE2001} and Gaussian approximation~\cite{GA2001} have been widely used for accurately estimating belief propagation (BP) decoding performance by tracking the evolution of message statistics during the decoding process.

\begin{figure}[t]
    \centering
    \includegraphics[width=\columnwidth]{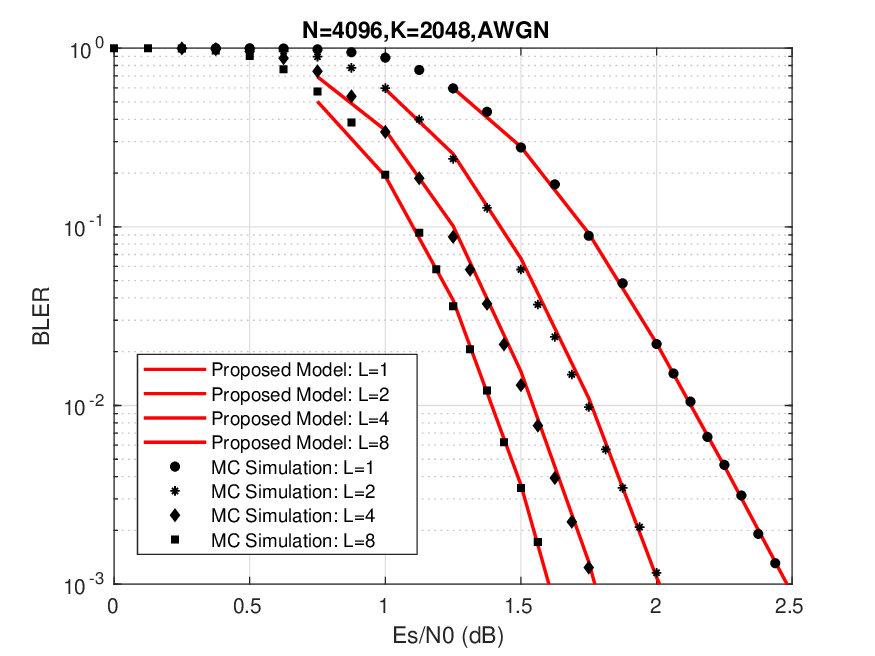}
    \caption{Analytical BLER predictions versus empirical Monte Carlo results for polar code ($N=4096, K=2048$) under various list sizes $L$.}
    \label{fig:performance}
\end{figure}

For polar codes~\cite{arikan2009}, CRC-aided successive cancellation list (CA-SCL) decoding~\cite{tal2015list, niu2013improved} remains the standard algorithm in many practical communication systems. However, a comparable analytical framework for predicting its finite-length performance has yet to be established. Mathematically modeling the highly non-linear path-pruning mechanism of CA-SCL presents two challenges:
\begin{enumerate}
\item \textbf{Modeling list-path management}: Modeling the list-management process is mathematically challenging because of the path splitting, path-metric sorting, and pruning operations that determine whether the correct path remains in the list or is discarded. Unlike analyses based mainly on density evolution or Gaussian approximation, which track the marginal probability distributions of messages or log-likelihood ratios, SCL list management involves order statistics and nonlinear selection operations. These features are not easily incorporated into standard probabilistic decoding-analysis frameworks.

\item \textbf{Correlation among decoding paths and error events}: The $L$ candidate paths in the list are generally correlated, since many of them share common prefixes and may be sibling branches originating from the same parent path. In addition, bit-decision error events are correlated because of the sequential nature of successive-cancellation decoding. These correlations further complicate the analysis of CA-SCL decoding.

\end{enumerate}
Consequently, predicting the exact performance of CA-SCL remains difficult, forcing researchers to rely heavily on computationally expensive Monte Carlo simulations for
performance assessment.

To bridge this gap, we propose a path-survival analytical framework for finite-length CA-SCL decoding. The core idea is straightforward: we model list decoding failure as a series
of distinct ``decoding crises'' rather than a gradual metric decline. This perspective allows us to decouple the complex path-pruning process into two tractable parameters: the expected
number of crises per block ($r$, which captures intrinsic code and channel characteristics and is list-independent), and the probability of surviving a single crisis ($p_L$, which captures the list gain and depends on the list size $L$). Consequently, the overall block error rate (BLER) can be analytically predicted for different list sizes using these parameters, avoiding the need for exhaustive list-specific simulations for each configuration. As illustrated in Fig.~\ref{fig:performance}, the proposed analytical curves align closely with Monte Carlo simulations across different list sizes under a wide range of signal-to-noise ratios (SNRs).

The remainder of this paper is organized as follows. Section~\ref{sec:scl_path_rank} reviews SCL decoding and the candidate path competition mechanism. Section~\ref{sec:psmodel} formulates the proposed path-survival model and derives a BLER prediction formula under this model. Section~\ref{sec:NumericalResults} presents extensive numerical results to prove the effectiveness of the proposed framework. Section~\ref{sec:discussion} discusses the relationship with existing works. Section~\ref{sec:Conclusion} concludes the paper.

\section{System Model and Path Rank Analysis}
\label{sec:scl_path_rank}

\subsection{Fundamentals of CA-SCL Decoding}
\label{subsec:scl_decoding}
SCL decoding simultaneously develops $L$ candidate paths, and tracks their likelihood through path metrics. At each bit index $i = 1, \dots, N$, the decoder performs three sequential operations:

\begin{enumerate}
    \item \textbf{Path splitting/extension:} If $u_i$ is an information bit, each active path splits into two branches ($\hat{u}_i \in \{0, 1\}$). If $u_i$ is a frozen bit, the path simply extends using the predefined frozen value (typically 0).

    \item \textbf{Path metric update:} The path metric (PM) is updated using the log-likelihood ratio (LLR) $L_i$. For a path extending with bit $\hat{u}_i$, the update is:
    \begin{equation}
    \mathrm{PM}_i = \mathrm{PM}_{i-1} +
    \begin{cases}
    0, & \text{if } \hat{u}_i = \hat{u}_i^{HD}; \\
    |L_i|, & \text{if } \hat{u}_i \neq \hat{u}_i^{HD};
    \end{cases}
    \end{equation}
    where $\mathrm{PM}_{i-1}$ is the path metric before decoding bit $u_i$, $L_i$ is the LLR for bit $u_i$, and $\hat{u}_i^{HD}$ denotes the hard decision based on the LLR, defined as:
    \begin{equation}
    \hat{u}_i^{HD} =
    \begin{cases}
    0, & \text{if } L_i \geq 0; \\
    1, & \text{if } L_i < 0.
    \end{cases}
    \end{equation}
    A smaller PM indicates a path with higher likelihood.

    \item \textbf{Path pruning:} The decoder sorts all generated candidates, retains the $L$ paths with the smallest metrics, and discards the rest.
\end{enumerate}

After all $N$ steps, CA-SCL outputs the path with the minimum metric among those that pass the CRC check. Because of the intermediate pruning operations, the correct path's survival depends entirely on its metric ranking relative to competing candidates. Unlike standard SC decoding, SCL tolerates temporary metric penalties. As long as the correct path remains within the top $L$ candidates at every decoding step, it survives. This relative ranking mechanism directly motivates the path-survival analysis detailed in the following subsection.

\subsection{Statistical Characterization of Path Rank Dynamics}
\label{subsec:rank_dynamics}
To analyze SCL performance, we track the metric rank of the correct path, denoted as $R_{\mathrm{true}}$, throughout the decoding process. At each step, paths are sorted by their metrics before the pruning operation. The correct path survives as long as $R_{\mathrm{true}} \leq L$, where a rank of $1$ indicates the best metric.

The trajectory of $R_{\mathrm{true}}$ depends fundamentally on the bit type: information bits induce path splitting, while frozen bits do not. This structural difference creates two opposing dynamics: \emph{rank degradation} at information bits and \emph{rank recovery} at frozen bits.

\begin{figure}[t]
\centering
\includegraphics[width=\columnwidth]{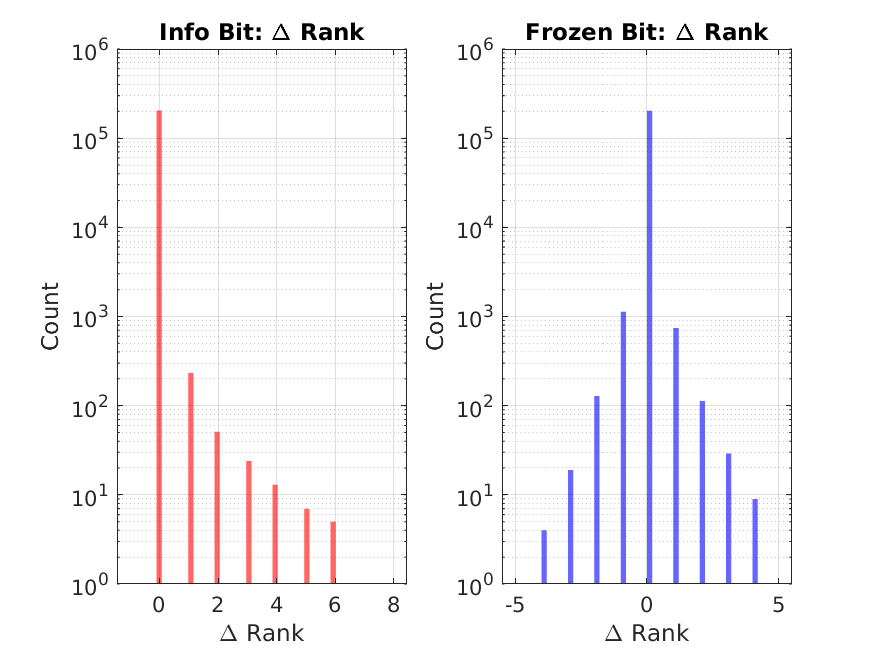}
\caption{Distribution of correct path rank changes ($\Delta \mathrm{Rank}$) during SCL ($L=8$) decoding. (a) Information bits; (b) Frozen bits. Simulation parameters: $N=4096$, $K=2048$, AWGN channel at $E_s/N_0 = 1.0~\mathrm{dB}$. Positive $\Delta \mathrm{Rank}$ indicates rank degradation, while negative $\Delta \mathrm{Rank}$ indicates rank recovery.}
\label{fig:rank_dynamics}
\end{figure}

\begin{itemize}
    \item \textbf{Rank degradation at information bits:} At information bits, each surviving path splits into two. If the local LLR points to the incorrect bit, the resulting erroneous branch acquires a smaller (better) metric than the correct one. Unlike SC decoding, this does not cause immediate failure; instead, it introduces a strong competitor into the list. After path-metric sorting, the correct path's ranking will drop.

    \item \textbf{Rank recovery at frozen bits:} At frozen bits, paths do not split. However, the path metric update incorporates LLRs based on prior decisions ($\hat{u}_1^{i-1}$). The correct path relies on true priors, typically incurring only a marginal metric penalty. In contrast, incorrect paths propagate previous errors, distorting their LLRs and incurring heavy metric penalties. This disproportionately punishes erroneous competitors, pushing them down the sorted list and allowing the correct path to recover its rank.
\end{itemize}

Fig.~\ref{fig:rank_dynamics} confirms the aforementioned rank drop and recovery behavior, where information bits typically produce $\Delta \mathrm{Rank} \geq 0$, while frozen bits are more likely to yield $\Delta \mathrm{Rank} < 0$. Most importantly, we observe the following phenomenon:
\begin{itemize}
    \item For more than $98\%$ of bit positions, the rank remains unchanged, i.e., $\Delta \mathrm{Rank} = 0$. This indicates that rank changes are highly concentrated: list failures are driven by sparse but severe error events, rather than by continuous and gradual degradation.
\end{itemize}

\begin{figure}[t]
\centering
\includegraphics[width=\columnwidth]{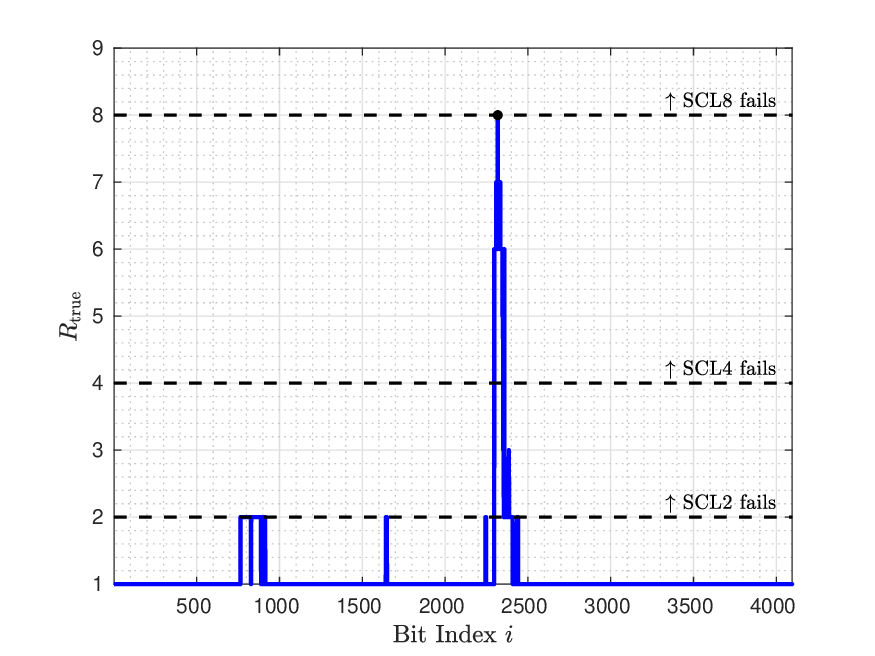}
\caption{An empirical trajectory of the correct path's rank during an SCL decoding process ($N=4096, L=8$).}
\label{fig:rank_trajectory}
\end{figure}

Fig.~\ref{fig:rank_trajectory} illustrates the rank trajectory of a correct path. The path remains at rank $1$ for most of the decoding process but is occasionally disrupted by sparse yet severe bit-error events. For instance, a severe crisis event sharply degrades its rank to $8$, bringing the correct path to the brink of elimination. However, the subsequent frozen bits strongly penalize competing erroneous paths and quickly restore the correct path back to rank $1$. This behavior confirms the cumulative yet recoverable nature of path competition.

These sparse degradation-recovery cycles form the basis of our analytical model. We formally define them as follows:

\begin{definition}[Decoding Crisis]
A decoding crisis is a localized event comprising two phases: (i) a degradation phase, where information-bit errors introduce strong competitors and drop the correct path's rank; and (ii) a recovery phase, where subsequent frozen bits penalize these competitors, restoring the correct path to rank 1. The crisis ends when the rank stabilizes or the block terminates.
\end{definition}

Because the rank recovery mechanism reliably resets the competition state ($R_{\mathrm{true}} \to 1$) after each crisis, distinct crises can be modeled as conditionally independent events. This critical insight provides the foundation for the tractable analytical framework detailed in Section~\ref{sec:psmodel}.

\section{Path-Survival Model for SCL Decoding}
\label{sec:psmodel}
As observed in Section~\ref{subsec:rank_dynamics}, list decoding failure is not a continuous decline, but rather the result of distinct, independent decoding crises. This observation motivates a path-survival view of SCL decoding: conditioned on the occurrence of several crises in one block, the correct path can remain in the final list only if it survives all of them.

Driven by this insight, we model the core determinants of SCL decoding performance with two decoupled parameters:
\begin{enumerate}
    \item $r$: the \textbf{expected crisis count}, denoting the average number of such crises per block. Since $r$ captures the inherent decoding difficulty induced by channel noise, it is independent of the list size $L$, depending exclusively on the channel conditions and the code structure.
    \item $p_L$: the \textbf{crisis-survival probability under SCL-$L$}, representing the likelihood that the correct path survives a single decoding crisis. It naturally depends on $L$; specifically, $p_L$ increases monotonically as the list size $L$ grows.
\end{enumerate}

This decoupling offers an analytical advantage as follows: $r$ can be extracted independently of the list decoding process. Once $r$ is determined, the BLER for any target list size $L$ can be analytically predicted by simply incorporating the corresponding $p_L$, thereby circumventing the need for exhaustive Monte Carlo simulations for each $L$.

The remainder of this section derives the two model parameters. Section~\ref{subsec:r_estimation} estimates the expected crisis count $r$, and Section~\ref{subsec:pl_derivation} derives the list-dependent survival probability $p_L$. Section~\ref{subsec:bler_prediction} then combines these parameters to obtain the BLER prediction formula under the proposed path-survival model.

\subsection{Expected Crisis Count $r$}
\label{subsec:r_estimation}
To estimate the expected decoding crisis count $r$, we propose a statistical approach based on the total bit errors $E_{\text{total}}$ under genie-aided successive cancellation (SC) decoding. Specifically, $E_{\text{total}}$ is defined as the sum of $K$ indicator variables, $E_{\text{total}} = \sum_{i=1}^K I_i$, where $I_i=1$ if the $i$-th information bit is incorrectly decoded under genie-aided SC.

\begin{figure}[t]
\centering
    \includegraphics[trim=0mm 60mm 0mm 60mm, clip, width=\columnwidth]{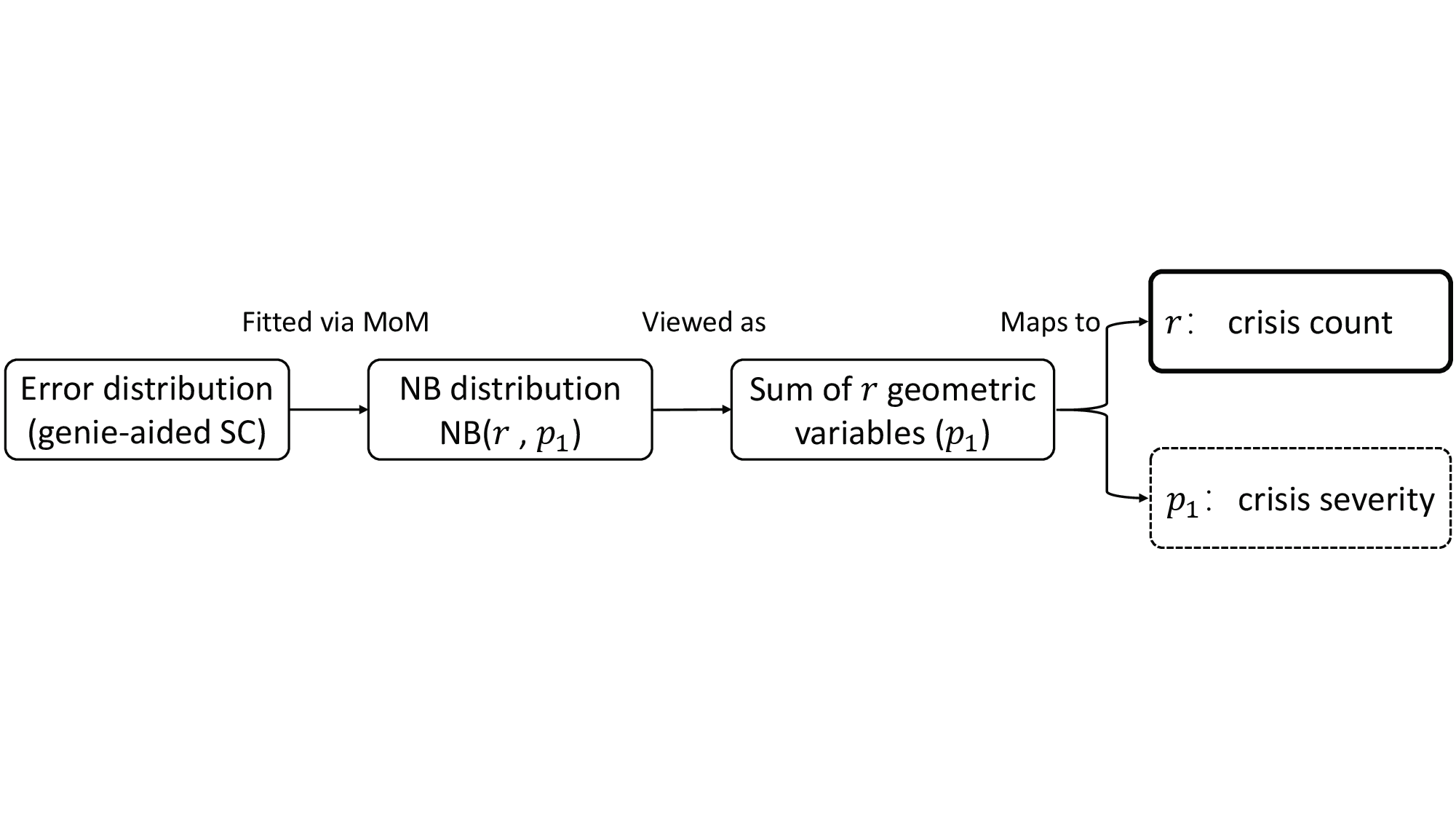}
\caption{Conceptual mapping from the empirical distribution of total bit errors ($E_{\text{total}}$) to the expected crisis count ($r$). The $E_{\text{total}}$ distribution under genie-aided SC decoding is fitted to a Negative Binomial (NB) model via the method of moments. Viewing the NB distribution as a sum of $r$ i.i.d. geometric variables reveals that the shape parameter $r$ directly represents the expected number of decoding crises.}
\label{fig:rmodel}
\end{figure}

\textbf{$E_{\text{total}}$ approximately follows NB distribution.} As illustrated in Fig.~\ref{fig:rmodel}, our core idea is to fit the empirical distribution of $E_{\text{total}}$ to a Negative Binomial (NB) distribution, where the shape parameter corresponds to $r$. The reason is as follows: in genie-aided SC decoding, although error propagation is completely eliminated, the error indicators $\{I_i\}_{i=1}^K$ still exhibit weak correlations due to their shared dependence on the same $N$ channel observations. Based on Stein's method~\cite{chen2010}, the NB distribution provides an accurate approximation for sums of such weakly correlated, integer-valued random variables~\cite{kumar2021, vellaisamy2013, cekanavicius2013}. Therefore, the NB distribution is well-suited for modeling $E_{\text{total}}$, which aligns with the findings of Liu et al.~\cite{liu2026polar}.
Conversely, while the standard Poisson distribution provides an accurate approximation for sums of \textit{independent} Bernoulli variables, these inherent weak correlations render the Poisson approximation~\cite{barbour1992} inapplicable.

\textbf{Interpretation of NB Parameters.} The choice of the NB model offers an interpretation of $r$. Mathematically, an NB random variable can be decomposed into the sum of $r$ independent and identically distributed (i.i.d.) geometric random variables. In our context, we associate each constituent geometric variable with a distinct decoding crisis. Consequently:
\begin{itemize}
  \item The shape parameter $r$ quantifies the expected number of decoding crises. A larger $r$ indicates more frequent crisis occurrences.
  \item The success probability $p_1$ denotes the termination probability of an individual decoding crisis. A larger $p_1$ implies faster crisis termination, corresponding to less severe rank degradation.
  \end{itemize}
Note that $r$ represents a statistical expectation and may take non-integer values.

\textbf{Parameter Estimation.} We employ the Method of Moments (MoM) to extract these parameters. By matching the theoretical mean $\mu$ and variance $\sigma^2$ of the NB distribution to the empirical moments of $E_{\text{total}}$, we obtain:
\begin{equation}
\mu = \frac{r(1-p_1)}{p_1}, \quad \sigma^2 = \frac{r(1-p_1)}{p_1^2}.
\end{equation}
Solving this system yields the empirical parameters:
\begin{equation}
r_{\text{fit}} = \frac{\mu^2}{\sigma^2 - \mu}, \quad p_{1,\text{fit}} = \frac{\mu}{\sigma^2}.
\label{eq:moment_matching}
\end{equation}

To apply Eq.~\eqref{eq:moment_matching}, explicit expressions for $\mu$ and $\sigma^2$ are required. These vary depending on the channel model:
\begin{itemize}
    \item \textbf{Binary Erasure Channel (BEC):} Exact closed-form expressions are available. {For the BEC, polarized sub-channels remain BECs~\cite{arikan2009}.} {The mean $\mu$ is derived from the Bhattacharyya parameters $Z(W_N^{(i)})$, which are calculated via the standard recursion $(p,p) \rightarrow (2p - p^2, p^2)$.} {To capture sub-channel correlations, the variance $\sigma^2$ incorporates the covariance matrix $C_N$.} {The matrix elements $C_N(i,j) \triangleq \text{Cov}[\mathcal{E}_i, \mathcal{E}_j]$ are computed recursively from $C_{N/2}$~\cite[Eq.~(19)]{liu2026polar}.} {The initial condition is $C_1 = p(1 - p)$.} The resultant moments are~\cite[Theorem 3]{liu2026polar}:
    \begin{equation}
    \mu = \frac{1}{2} \sum_{i \in \mathcal{I}} Z(W_N^{(i)}), \quad \sigma^2 = \frac{1}{2}\mu + \frac{1}{4} \sum_{i, j \in \mathcal{I}} C_N(i,j). \label{eq:moments_bec}
    \end{equation}

    \item \textbf{AWGN/BSC Channels:} For continuous or non-erasure discrete channels, the mean $\mu$ is computed analytically via density evolution~\cite{tal2015list} as $\mu = \sum_{i \in \mathcal{I}} P_e^{(i)}$. However, tracking the joint bit-error correlations across polarization steps to derive an exact variance $\sigma^2$ is mathematically prohibitive due to the complex dependency structure. Therefore, we adopt a hybrid approach where $\sigma^2$ is estimated empirically via Monte Carlo simulations.
\end{itemize}

\subsection{Crisis-Survival Probability $p_L$}
\label{subsec:pl_derivation}
To evaluate the survival probability $p_L$ of a single decoding crisis under list size $L$, we analyze the number of accumulated information-bit errors, $E$. As discussed in Section~\ref{subsec:r_estimation}, the distribution of $E$ is modeled as a geometric distribution:
\begin{equation}
    \Pr(E=k) = (1-p_1)^k p_1, \qquad k=0,1,2,\ldots .
    \label{eq:geo_dist}
\end{equation}

\begin{figure}[t]
    \centering
    \includegraphics[trim=25mm 78mm 70mm 15mm, clip, width=\columnwidth]{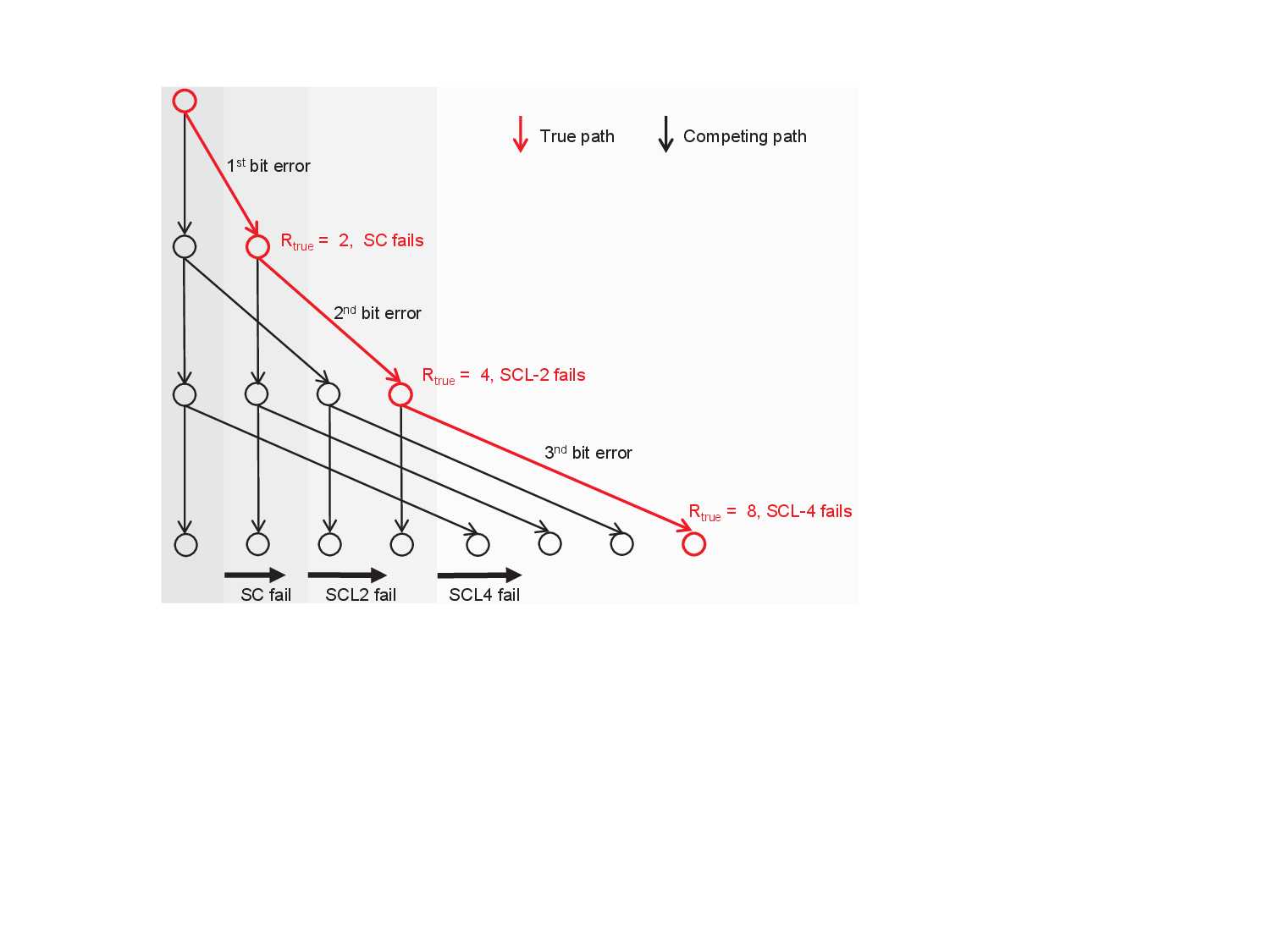}
    \caption{Illustration of the \emph{approximate} evolution of correct path rank degradation during a decoding crisis. As the number of information bit errors $k$ increases, the rank of the correct path grows exponentially ($R_{\mathrm{true}}(k) \approx 2^k$).}
    \label{fig:rank_growth}
\end{figure}

To quantify its impact on the survival probability $p_L$, we examine how $k$ errors degrade the ranking of the correct path. During a decoding crisis, when the correct path encounters an information-bit error, its PM degrades, and an erroneous competing path with a better PM is simultaneously introduced. As these surviving erroneous paths continue to split at subsequent information bits, they progressively generate additional paths with better PMs. Consequently, the total number of competing paths ranked ahead of the correct path grows exponentially. For the BEC, it has been proven in~\cite[Appendix A]{QuantizedPolar} that after $k$ such erasures, there exist exactly $2^k$ paths that satisfy the survival criteria, yielding a worst-case rank of $2^k$ for the correct path. For AWGN and BSC channels, we lack the rigorous analysis available for BECs, but similar behaviors can still be empirically observed. Motivated by this, as illustrated in Fig.~\ref{fig:rank_growth}, we approximate the evolution of the correct path's rank, denoted by $R_{\mathrm{true}}(k)$, after $k$ errors as follows:
\begin{equation}
    R_{\mathrm{true}}(k) \approx 2^k .
    \label{eq:rank_doubling}
\end{equation}

For SCL decoding, the correct path survives pruning if its rank does not exceed the list size $L$:
\begin{equation}
    R_{\mathrm{true}}(k) \leq L .
\end{equation}
Using~\eqref{eq:rank_doubling}, the survival condition becomes:
\begin{equation}
    2^k \leq L \quad \Longrightarrow \quad k \leq \left\lfloor \log_2 L \right\rfloor .
    \label{eq:survival_condition}
\end{equation}
Therefore, the crisis-survival probability $p_L$ is calculated as:
\begin{equation}
\begin{split}
    p_L &\approx \Pr\left(E \leq \left\lfloor \log_2 L \right\rfloor\right) \\
    &= \sum_{k=0}^{\left\lfloor \log_2 L \right\rfloor} (1-p_1)^k p_1 \\
    &= 1 - (1-p_1)^{\left\lfloor \log_2 L \right\rfloor + 1}.
\end{split}
\label{eq:pL_derivation}
\end{equation}
\begin{figure*}[!t]
    \centering
    \includegraphics[trim=35mm 15mm 35mm 10mm, clip, width=\textwidth]{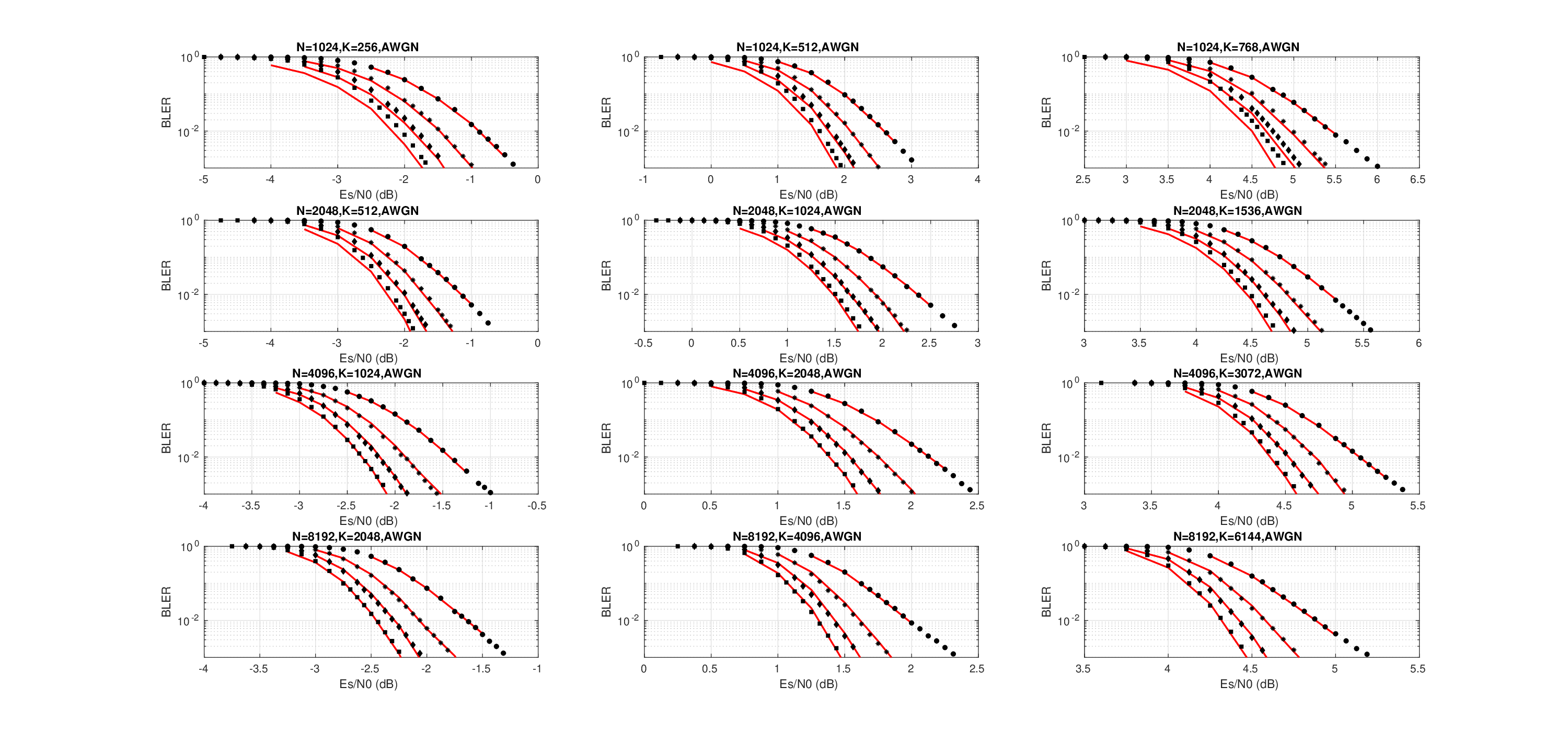}
    \caption{Performance over the AWGN Channels. \textbf{Solid lines} denote the proposed analytical predictions for $L \in \{1, 2, 4, 8\}$, while \textbf{dotted lines} correspond to the MC simulation benchmarks.}
    \label{fig:results_grid}
\end{figure*}

\begin{figure*}[!t]
    \centering
    \includegraphics[trim=35mm 15mm 35mm 10mm, clip, width=\textwidth]{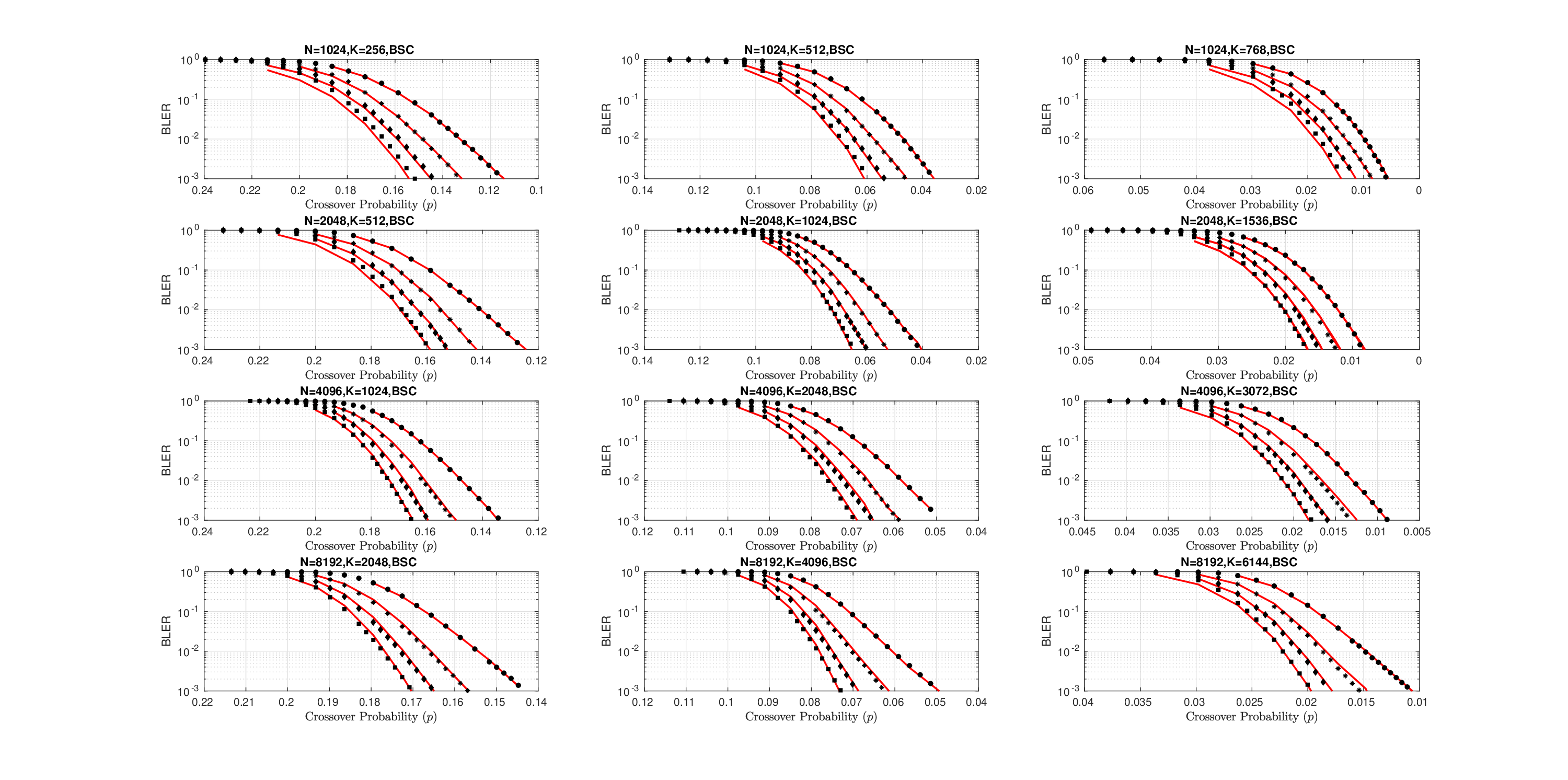}
    \caption{Performance over the BSC Channels. \textbf{Solid lines} denote the proposed analytical predictions for $L \in \{1, 2, 4, 8\}$, while \textbf{dotted lines} correspond to the MC simulation benchmarks.}
    \label{fig:results_grid_bsc}
\end{figure*}
Intuitively, this implies that every doubling of the list size $L$ increases the tolerable error threshold by one bit, thereby reducing the probability of path pruning within a crisis. The accuracy of this proposed approximation will be numerically validated in Section~\ref{sec:NumericalResults}.

\subsection{BLER Prediction Framework}
\label{subsec:bler_prediction}
With the intrinsic parameters $(r_{\text{fit}}, p_{1,\text{fit}})$ and the derived list-dependent survival probability $p_L$, we now synthesize the overall SCL BLER prediction model.

A key assumption underlying this synthesis is the statistical independence of distinct decoding crises. As observed in Section~\ref{subsec:rank_dynamics}, the rank recovery mechanism at frozen bits effectively resets the path competition state ($R_{\mathrm{true}} \to 1$) after a crisis resolves. This reset suggests that consecutive crises are weakly correlated, allowing us to approximate their survival outcomes as independent events. Consequently, the probability that the correct path successfully survives \textit{all} crises within a decoding block is modeled as the product of their individual survival probabilities, $p_L^r$.

By substituting the parameters, the predicted BLER for list size $L$ is formulated as:
\begin{equation}
    P_{\text{SCL}}^e(L) = 1 - p_L^{r_{\text{fit}}}.
    \label{eq:bler_final}
\end{equation}

The complete workflow of the analytical framework for SCL performance prediction is summarized in Algorithm~\ref{alg:analytical_prediction}. It operates in two stages: (1) acquiring the \emph{$L$-independent} empirical parameters, $r_{\text{fit}}$ and $p_{1,\text{fit}}$, via the method of moments, and (2) computing the \emph{$L$-dependent} path survival probability, $p_L$, to predict the final BLER of the SCL-$L$ decoder.

\begin{algorithm}[t]
    \caption{Analytical SCL Performance Prediction}
    \label{alg:analytical_prediction}
    \begin{algorithmic}[1]
        \Require Intrinsic parameters $(\mu, \sigma^2)$, Target List Size $L$
        \Ensure Predicted BLER $P_{\text{SCL}}^e(L)$
        \State Extract intrinsic parameters $r_{\text{fit}}$ and $p_{1,\text{fit}}$ using (\ref{eq:moment_matching})
        \State Compute path survival probability $p_L$ using (\ref{eq:pL_derivation})
        \State \Return $P_{\text{SCL}}^e(L) = 1 - p_L^{r_{\text{fit}}}$
    \end{algorithmic}
\end{algorithm}

\section{Numerical Results}
\label{sec:NumericalResults}

\subsection{Validation of the Analytical Model}
\label{subsec:model_validation}

To evaluate the proposed analytical framework, we compare the analytical predictions with Monte Carlo (MC) simulations, as presented in Figs.~\ref{fig:results_grid} and~\ref{fig:results_grid_bsc}. The evaluation covers both AWGN and BSC channels across various code lengths $N \in \{1024, 2048, 4096, 8192\}$ and rates $R \in \{1/4, 1/2, 3/4\}$. The polar codes are constructed using the Gaussian Approximation (GA) method and are concatenated with a standard $16$-bit CRC.

The results demonstrate close agreement between the analytical predictions and the empirical MC simulation results across a wide range of configurations. This alignment confirms that the proposed framework accurately captures the performance gain as the list size $L$ increases. More importantly, it validates our core analytical premise: the $L$-independent parameters $r$, derived from genie-aided statistics, can robustly predict the $L$-dependent SCL decoding performance. This is particularly evident for moderate to long block lengths ($N \ge 2048$), where the analytical curves and simulation markers almost perfectly overlap.

In the regime of short block lengths (e.g., $N=1024$), we observe that the analytical prediction slightly underestimates the simulation results at larger list sizes ($L=8$). This deviation arises because shorter codes possess fewer frozen bits between consecutive information bits. Consequently, the rank recovery mechanism lacks sufficient separation to fully reset the path competition state ($R_{\mathrm{true}} \to 1$) before a subsequent decoding crisis is triggered. This introduces mild inter-crisis correlations, which slightly violate the independence assumption of the model. Nevertheless, the expected SNR gap between the predicted performance and the simulated curves remains strictly within $0.1$~dB.

\begin{figure}[!t]
\centering
\includegraphics[width=\columnwidth]{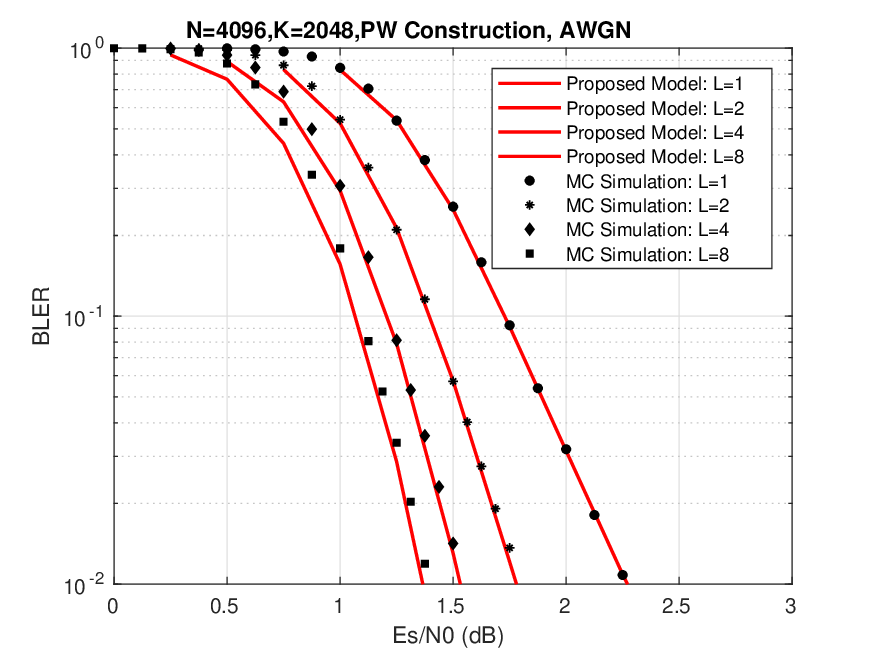}
\caption{Analytical BLER predictions match closely with Monte Carlo simulation results for a polar code constructed based on a PW sequence ($N=4096, K=2048$) under various list sizes $L$.}
\label{fig:4096PW}
\end{figure}

To assess the generalizability of the proposed model, we also evaluate its predictions for polar codes constructed under different information set configurations. Specifically, we consider the Polarization Weight (PW) sequence~\cite{PW}, an efficient offline construction method whose principles closely align with the reliability-sequence-based designs adopted in 5G New Radio (NR) standards. As depicted in Fig.~\ref{fig:4096PW}, the analytical predictions maintain a close alignment with the Monte Carlo simulations. This consistency substantiates the robustness of the proposed framework across different information set selections.

Collectively, these results validate the framework's capability to deliver accurate finite-length performance predictions across diverse code rates, block lengths, list sizes, and information set configurations.

\subsection{Interpretation of Model Parameters}
\begin{figure}[!t]
\centering
\includegraphics[width=\columnwidth]{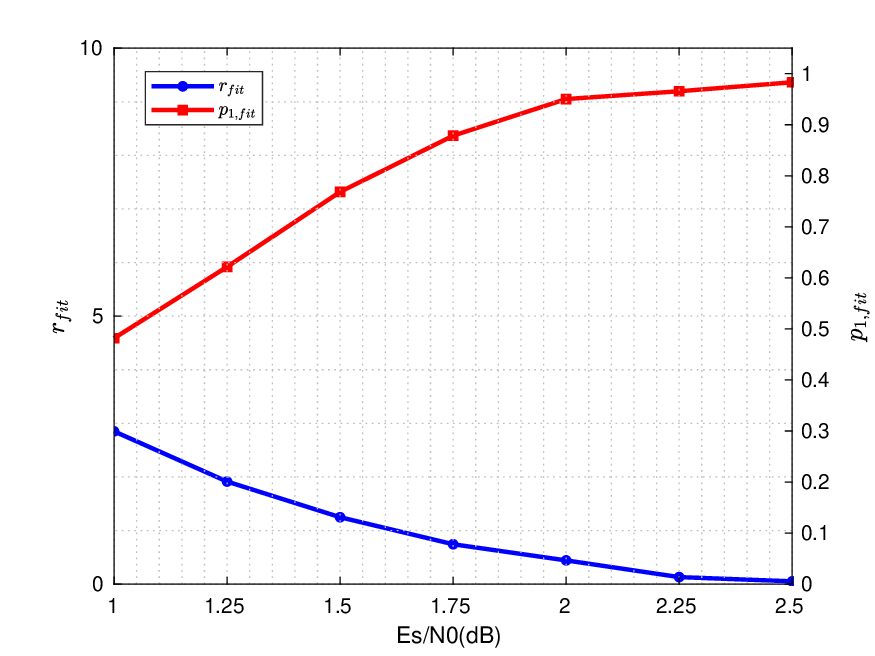}
\caption{Model parameters versus signal-to-noise ratio ($N=4096, K=2048$). The blue curve represents the expected crisis count $r_{\text{fit}}$ (left y-axis), while the red curve denotes the crisis termination probability $p_{1,\text{fit}}$ (right y-axis).}
\label{fig:fit_params_vs_esn0}
\end{figure}

We examine the behavior of model parameters across different SNRs to gain insights into the error distribution characteristics of the correct path. Recall that $r_{\text{fit}}$ represents the expected number of decoding crises, while $p_{1,\text{fit}}$ denotes the probability of error termination within each crisis. As shown in Fig.~\ref{fig:fit_params_vs_esn0}, $r_{\text{fit}}$ decreases monotonically, while $p_{1,\text{fit}}$ increases monotonically as channel conditions improve. These trends reflect the evolution of the decoding dynamics:
\begin{itemize}
    \item The decrease in $r_{\text{fit}}$ indicates that the decoder encounters fewer decoding crises at higher SNRs, as the channel becomes more reliable.
    \item The increase in $p_{1,\text{fit}}$ demonstrates that each crisis terminates more quickly, meaning errors are less likely to propagate consecutively.
\end{itemize}

These monotonic trends confirm that the model effectively captures the underlying error behavior across SNR regimes. The interpretation is as follows:
\begin{itemize}
    \item \textbf{High SNRs:} The combination of low $r_{\text{fit}}$ (few crises) and high $p_{1,\text{fit}}$ (short crisis duration) implies that errors manifest as sporadic and isolated events. Since crises are both infrequent and quickly resolved, errors do not have sufficient opportunity to accumulate or propagate.
    \item \textbf{Low SNRs:} Conversely, the combination of high $r_{\text{fit}}$ (frequent crises) and low $p_{1,\text{fit}}$ (long crisis duration) suggests that errors tend to propagate as dense, clustered bursts. With many crises occurring and each persisting for longer durations, errors accumulate in clusters before the correct path can recover its rank.
\end{itemize}
This transition from clustered bursts at low SNRs to isolated events at high SNRs aligns with the intuitive understanding of polar code decoding behavior, validating the interpretability of the proposed model.

\section{Relationship with Existing Works}
\label{sec:discussion}
In this section, we review existing analytical methods for polar codes, and discuss their relationship with our proposed path-survival model.

\begin{itemize}
\item \textbf{Finite-length performance prediction:} This line of work aims to calculate or bound the decoding performance for specific, finite block lengths. For instance, Tal and Vardy~\cite{tal2013construct} employ DE to calculate the error probabilities of individual bit-channels, which are then aggregated via the union bound to provide an accurate estimation of SC decoding performance in the high-SNR regime. Recently, Liu et al.~\cite{liu2026polar} explore statistical modeling approaches and demonstrate that error counts under genie-aided SC decoding follow a negative binomial distribution, enabling precise SC performance prediction over a wider SNR range.

    \textit{Relationship to our work:} While these methods provide performance bounds for SC decoding, they cannot be trivially extended to CA-SCL decoding. Our path-survival framework bridges this gap by capturing the rank recovery mechanism enabled by frozen bits. To mathematically quantify this process, we build upon existing tools; specifically, similar to~\cite{tal2013construct}, we leverage DE to estimate the key parameter $\mu$ required for our model. Notably, when the list size $L=1$, our framework reduces to the SC decoding case and aligns with Liu et al.'s statistical model~\cite{liu2026polar}.

\item \textbf{Asymptotic analysis:} These works study the scaling laws and fundamental limits as the block length $N \to \infty$. Ar{\i}kan and Telatar~\cite{arikan2009rate} established the error exponent for SC decoding. Hassani et al.~\cite{hassani2014finite} derived finite-length scaling bounds to explain how block length and gap-to-capacity impact reliability, which Mondelli et al.~\cite{mondelli2015scaling, mondelli2016unified} later extended to list decoding.

    \textit{Relationship to our work:} While scaling bounds provide theoretical insights into asymptotic behavior ($N \to \infty$), they do not yield explicit BLER predictions for practical finite block lengths (e.g., $N = 1024 \to 8192$). Notably, in~\cite{mondelli2015scaling}, the scaling factor under SCL decoding is shown to coincide with that of SC decoding, failing to capture the extra gain of list decoding in the finite blocklength regime. Moreover, such theoretical analyses typically assume optimal information bit selection for SC decoding, thus failing to distinguish different code constructions. Our path-survival framework complements these theoretical bounds, enabling more precise finite-length performance estimation for specific codes.
\end{itemize}

\section{Conclusion}
\label{sec:Conclusion}
This paper presents an analytical path-survival model to evaluate the finite-length performance of CA-SCL decoding. By characterizing the decoding process as a sequence of degradation-recovery crises, we establish an explicit mapping from genie-aided SC error statistics to SCL survival probabilities. This approach enables accurate BLER prediction for SCL decoding without the need of list-specific Monte Carlo simulations. Extensive numerical results over BSC and AWGN channels validate the accuracy of the proposed framework across various code lengths, rates, and list sizes. By capturing the rank recovery mechanism enabled by frozen bits, this framework provides a practical analytical tool for evaluating and understanding SCL decoding behaviors.

\balance

\end{document}